# Hybridized/Coupled Multiple Resonances in Nacre

Seung Ho Choi and Young L. Kim[*]
*Weldon School of Biomedical Engineering, Purdue University*
*206 S. Martin Jischke Drive, West Lafayette, IN, 47907*
* youngkim@purdue.edu

We report that nacre (also known as mother-of-pearl), a wondrous nanocomposite found in nature, is a rich photonic nanomaterial allowing the experimental realization of collective excitation and light amplification via coupled states. Localized modes in three-dimensional complex media are typically isolated in frequency and space. However, multiple local resonances can be hybridized in multilayered nanostructures of nacre so that the effective cavity size for efficient disordered resonators is scaled up. Localized modes in hybridized states in nacre are overlapped in frequency with similar shapes in space, thus being collectively excited and synergistically amplified. These hybridized states boost light amplification, leading to stable and regular multimode lasing at low excitation energy. The simplicity of ameliorating disordered resonators by mimicking nacre can further serve as platforms for developing cost-effective photonic systems and provide materials for fundamental research on complex media.

PACS numbers: 83.80.Mc 05.45.Ac 42.55.Zz 78.55.Qr

## I. INTRODUCTION

Interactions of electromagnetic waves with irregular and open structures are commonly considered to be inefficient, although disordered systems can be effectively used to identify disorder-induced resonances for enhanced confinement and transport.[1-4] Light amplification in such media[5-9] requires high excitation power and has low energy conversion efficiency, hampering its practical and widespread applications. Electrical excitation schemes have also been used to avoid high-power short pulse optical excitation.[10-12] To overcome the intrinsic limitations of disordered systems, simultaneous utilization of multiple resonances can be an effective approach for efficiently exciting disordered resonators. Indeed, multiple resonances can be coupled to form modes overlapped both in space and frequency (also known as necklace states): A collection of these modes can be hybridized into a long chain with multiple peaks in intensity. Hybridized/coupled states have often been studied in two regimes defined by the degree of spectral mode overlap (i.e. Thouless number or fundamental localization parameter): mode spacing $\Delta\omega <$ linewidths $\delta\omega$ and $\Delta\omega \geq \delta\omega$.[13-19]

Natural nacre can serve as intriguing low-dimensional photonic nanostructures to capitalize on hybridized multiple resonances for enhanced light-matter interactions. From a mechanical standpoint, as nanocomposites found in the inside layers of abalone shells, nacre has received considerable attention:[20-25] The mechanical and structural properties of nacre of abalone shells have been intensively studied to better understand how they deform and fracture in terms of plasticity and toughness. Indeed, nacre has provided valuable clues for synthesizing materials that mimic their excellent mechanical and structural properties.[26-28] From an optical standpoint, nacre has been studied to better understand the unique colors in terms of diffraction from fine-scale grating structures on the shell surface and interference from the inner nacreous layers.[29-31] In particular, photonic band states, which imply perfect ordered nanostructures, are speculated to play an important role in the distinct colors of abalone shells and pearls.[32,33] However, these unique nanostructures of nacre have not yet been exploited as optical resonators.

We report that hybridized/coupled multiple resonances can be possible underlying states for boosting light amplification and lasing in nacre, because nacre is a partially disordered nanostructure with an extremely large number of parallel layers. Although other previously identified natural/biological materials are often inappropriate for direct use in photonics,[34] nacre serves as an immediately exploitable nanocomposite to study multiple resonances in low dimension. In particular, the internal field intensity of coupled states in layered structures was investigated using large scale microwave experiments while nonlinear effects could not be studied.[16,17] In photonics, mechanisms of necklace/coupled states, by which light can be effectively coupled into lasing states, have not yet been studied, in part because photonic structures realizing and visualizing hybridized/coupled states are rare in nature and challenging to synthesize. In this study, we investigate hybridized multiple resonances and visualize their local energy density distributions, in which gain media make such resonances readily detectable. First, we quantify the physical and nanostructural properties of nacre of green abalone shells. Second, using numerical simulations based on the transfer matrix method, we study the role of hybridized multiple resonances in light transport and amplification. Third, we conduct photoluminescence and lasing experiments using nacre.

## II. NANOSTRUCTURES AND PHOTONIC PROPERTIES OF NACRE

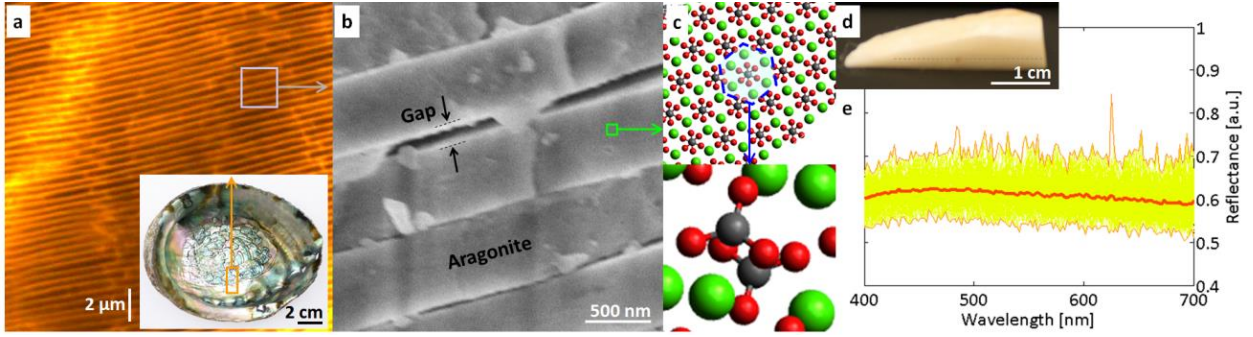

**FIG. 1.** (Color online) Nanostructures and photonic properties of *Haliotis fulgens* (green abalone). (**a**) Confocal fluorescence microscopy image of nacre infiltrated with rhodamine 6G (Rh6G) after deproteinization. Inset: Photograph of the original abalone shell. (**b**) Scanning electron microscopy (SEM) image of a brick-mortar nanostructure in nacre. (**c**) The molecular architecture of aragonite viewed from the C axis (top) and off the C axis (bottom). (**d**) Photograph of deproteinized nacre. (**e**) Reflectance spectra (spectral resolution = 2 nm) from deproteinized nacre, averaged from ~ 100 spectra.

We prepared nacre specimens from shells of *Haliotis fulgens* (green abalone and inset of Fig. 1(a)): Large strips from the shell were cut using a precision saw. The calcitic and growth layers were removed using a fine grinder. We also removed the organic materials (i.e. protein and chitin) that fill the gap of ceramic (i.e. aragonite) layers using sodium hypochlorite. In particular, the resulting gap allowed fluorescence dyes to easily smear inside nacre for fluorescence confocal microscopy (Fig. 1(a)).

Fig. 1(a) shows that nacre has inorganic ceramic (i.e. aragonite) tablets of 473 ± 50 (standard deviation, SD) nm arranged in layers. The ceramic tablets are separated by a gap of 40 ± 10 (SD) nm (Fig. 1(b)), which is occupied by organic materials (i.e. protein and chitin). Nacre contains an extremely large number of parallel layers. Nacre specimens with a thickness of 2 mm can contain up to ~ 4,000 alternating aragonite and gap layers. Aragonite has metastable pseudo-hexagonal crystal structures (Fig. 1(c)), resulting in an optically dense form of calcium carbonate ($CaCO_3$) polymorphs. Thus, nacre can be an ideal nanocomposite with alternating layers with low and high refractive indices (refractive index of aragonite $n_a$ ~ 1.65 − 1.68 and Fig. 2(b)). Moreover, the ceramic layers have a thickness comparable to the wavelength of the visible light. In this case, the phase can be randomized quickly as the wave propagates, leading to a short localization length.

Surprisingly, Fig. 1(d) reveals that nacre has a diffuse white color after deproteinization, although the bulk aragonite crystal is known to be optically translucent or transparent. The original captivating color of abalone nacre (inset of Fig. 1(a)) should be ascribed to both pigmentation of the organic materials and optical effects of the aragonite layers. The averaged reflectance spectrum from deproteinized thick nacre (Fig. 1(e)) is relatively flat in the visible wavelength range of $\lambda$ = 400 − 700 nm (assuming the corresponding perfect periodic structure, the bandstop is $\lambda$ ~ 555 − 569 nm): Reflectance spectra were obtained within a small angular cone of ± 2° in the backward direction[35] while deproteinized nacre was placed in a methanol solution (i.e. dye solvent). Indeed, the ratio of a SD to an averaged thickness of the aragonite- and gap-combined layer, which can be quantified as morphological disorder, is $\sqrt{50^2+10^2}/(473+40) \approx 0.1$. Thus, the unique color of nacre is significantly attributed to the disordered multilayered nanostructures.

## III. THEORETICAL CONSIDERATION OF HYBRIDIZED STATES IN NACRE

Given the highly multilayered nanostructures of nacre, we numerically investigated behavior of hybridized states assuming one-dimensional (1D) disordered cavities. For quasimodes and lasing modes in nacre's nanostructures, we used the transfer matrix method[35, 36] with complex and frequency-dependent refractive indices implementing linear gain,[37, 38] as shown in Fig. 2. By incorporating a spatially modulated frequency-dependent susceptibility $\chi_g(z, k = 2\pi/\lambda)$ in the refractive index of gap $n_g$, the homogeneous field amplification and the linear gain of dye molecules can be simulated such that

$$n_g(z,k) = n_r(z,k) + in_i(z,k) = \sqrt{n_{g(nr)}^2(z,k) + \chi_g(z,k)} \text{ and}$$

$$\chi_g(z,k) = \frac{A_m N_{ex}(z)}{k_a^2 - k^2 - ik\Delta k_a},$$

where $n_r$ and $n_i$ (< 0) are the real and imaginary parts of the refractive index, $n_{g(nr)}$ is the refractive index of the nonresonant background material, $z$ is the spatial coordinate, $A_m$ is a material-dependent constant, $N_{ex}$ is the density of excited atoms, $k_a$ is the atomic transition frequency (i.e. peak frequency of dye molecules), and $\Delta k_a$ is the spectral linewidth of the atomic resonance (i.e. spectral width of spontaneous emission of dye molecules). Both Re $\chi_g(k)$ and Im $\chi_g(k)$ are proportional to the density of excited atoms $N_{ex}$ determined by the excitation energy. It should be noted that this linear gain model is only valid at or below the threshold.

We accounted for the nacre's nanostructures using the structural parameters obtained from confocal microscopy and scanning electron microscopy (Fig. 1(a)&(b)). A similar level of morphological disorder was set by randomly varying the aragonite thickness and the gap distance in the ranges of $d_a = 473 \pm 50$ (SD) nm and $d_g = 40 \pm 10$ (SD) nm, respectively. The 1D nanostructures consisted of a total of 2,000 layers alternating aragonite and gap, corresponding to a system length of $L_T \sim 0.5$ mm. The wavelength-dependent refractive indices of aragonite and gap (methanol) were obtained from the experimental data[39, 40] and their dispersion curves are shown in Fig. 2(b). As low dye concentrations were used in later lasing experiments, the real part of the refractive index of dye solutions $n_g$ can be assumed to be close to that of the solvent itself (i.e. methanol).[41] Fig. 2(c) shows the real and imaginary parts of $\chi_g$ over the wavelength for the gain medium in the gap. For the solution of rhodamine 6G (Rh6G), $k_a = 11.38$ μm$^{-1}$ (551 nm) and $\Delta k_a = 0.72$ μm$^{-1}$ (35 nm) were obtained from the experimental fluorescence spectra. Using the optical and geometrical parameters closely mimicking the experimental conditions, a 1D localization length $\xi$ ($= -2L_T / <\ln T>$, where $<\ldots>$ stands for averaging over multiple realizations) was estimated that $\xi = 37.2$ μm. Thus, the system is in a localized regime $L_T \gg \xi$.

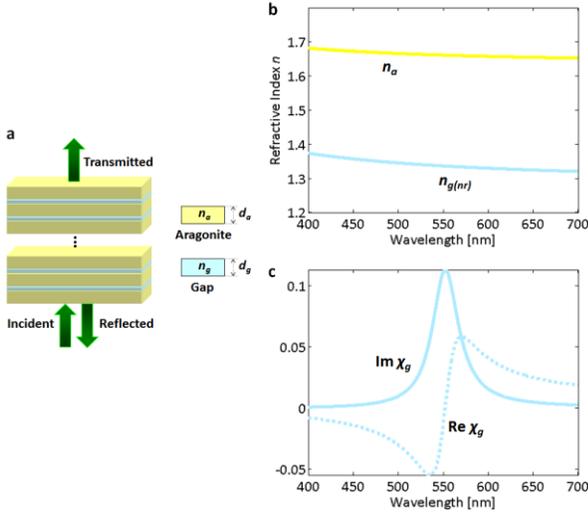

**FIG. 2.** (Color online) Numerical experiments of hybridized/coupled states and light amplification in 1D disordered systems. (**a**) Schematic diagram of the experimental configuration. (**b**) Dispersion curves of aragonite and gap layers. (**c**) Real and imaginary parts of susceptibility of gain $\chi_g$ ($\lambda$).

Fig. 3(a) shows calculated transmittance spectra $T$ and internal field intensity $I$ of resonance modes when gain is absent (quasimodes). As indicated by the representative cases (dotted boxes in Fig. 3(a)), we categorized the coupling status into isolated modes and hybridized states. Hybridized states are further divided into two groups: i) mode spacing $\Delta\omega <$ linewidths $\delta\omega$ and ii) $\Delta\omega \geq \delta\omega$. While both groups can be considered as coupled resonances with a different level of coupling, each can be identified using the accumulated phase shift[38] and the valleys between peaks in $\log_{10}(T)$[19] in addition to the fundamental localization parameters.

In particular, the transmission spectrum in the range of $\lambda = 551.5 - 552$ nm includes five narrow peaks close to one another. This group ($\Delta\omega \geq \delta\omega$) is the collection of localized modes (lower panel in Fig. 3(a)) which are overlapped in frequency and have similar field distributions in space.[16, 17] In this case, the relatively high valleys between the narrow adjacent peaks become obvious in $\log_{10}(T)$. Interestingly, when $\Delta\omega \geq \delta\omega$, the spatial field distributions are extended over the whole structure, consisting of multiple localized resonances in space, while individual $\delta\omega$ is still narrow. These states would have lifetimes shorter yet comparable to the single localized modes, because $\delta\omega$ of the adjacent peaks stays narrow while still having the very low valleys in the linear scale of $T$. However, when $\Delta\omega < \delta\omega$, the hybridized states would have much shorter lifetime because weak mode splitting occurs, resulting in a single broadened peak. Thus, in the nominally localized systems, the hybridized states with $\Delta\omega \geq \delta\omega$ are distinguishable from the hybridized states with $\Delta\omega < \delta\omega$. In this respect, the hybridized states with $\Delta\omega \geq \delta\omega$ are hereafter referred to as hybridized localized states.

Upon the introduction of the density of excited atoms $N_{ex} = 3 \times 10^{-3}$ which is spatially uniform in each gain region, the emission intensity $I_e$ of the hybridized localized state (within $551.5 - 552$ nm) is higher than those of the other hybridized states and isolated modes (Fig. 3(b)). As the gain increases, $I_e$ of the hybridized localized state is amplified together (Fig. 3(d)). To also compare with spatial images of intensity obtained using a microscopy imaging setup with finite bandwidth (Fig. 4(a)), the internal intensity is integrated such that $I(z) = \int I(\lambda, z) d\lambda$ over the range $\lambda = 550 - 554$ nm. When the internal field is averaged over the wavelength range, the passive hybridized localized state (without gain) is hidden (green curve in Fig. 3(c)). However, in the presence of gain (blue curve in Fig. 3(c)), the multiple peaks of the hybridized localized state dominantly emerge with an inter-resonance spacing of $41 \pm 19$ μm on the order of $\xi$, indicating a high coupling efficiency. In other words, gain allows the clear identification of the hybridized localized states formed from regularly spaced resonances in space.

To understand the light transport and amplification properties of the hybridized states, we analyzed 546 spectral peaks from 32 different configurations of 1D disordered cavities (Fig. 3(e)). The occurrence probability of all hybridized states (the number of the peaks in the hybridized states over the number of the total peaks) is similar to that of the isolated modes (inset of Fig. 3(e)). The overall tendency (Fig. 3(e)&(f)) is that the transmittance $T$ of the hybridized states (both $\Delta\omega < \delta\omega$ and $\Delta\omega \geq \delta\omega$) is $\sim 7.9$ times higher than that of the isolated modes, revealing that their main characteristic is enhanced transport or tunneling. In the hybridized localized states, the individual modes are overlapped in frequency with similar shapes spanning over the whole structure, thus being collectively excited. In addition, their spatially extended fields can utilize larger areas with gain. As a consequence of synergical amplification, $I_e$ of the hybridized localized states ($\Delta\omega \geq \delta\omega$) is $1.1 \times 10^3$ times higher than that of the isolated modes (Fig. 3(g)). $I_e$ of the hybridized localized states is twice higher than that of the hybridized states with $\Delta\omega < \delta\omega$. Interestingly, when $\Delta\omega < \delta\omega$, the hybridized states are still significantly amplified compared with the isolated modes (Fig. 3(g)), although their passive states appear to have the degenerate localized properties with the broadened $\delta\omega$.

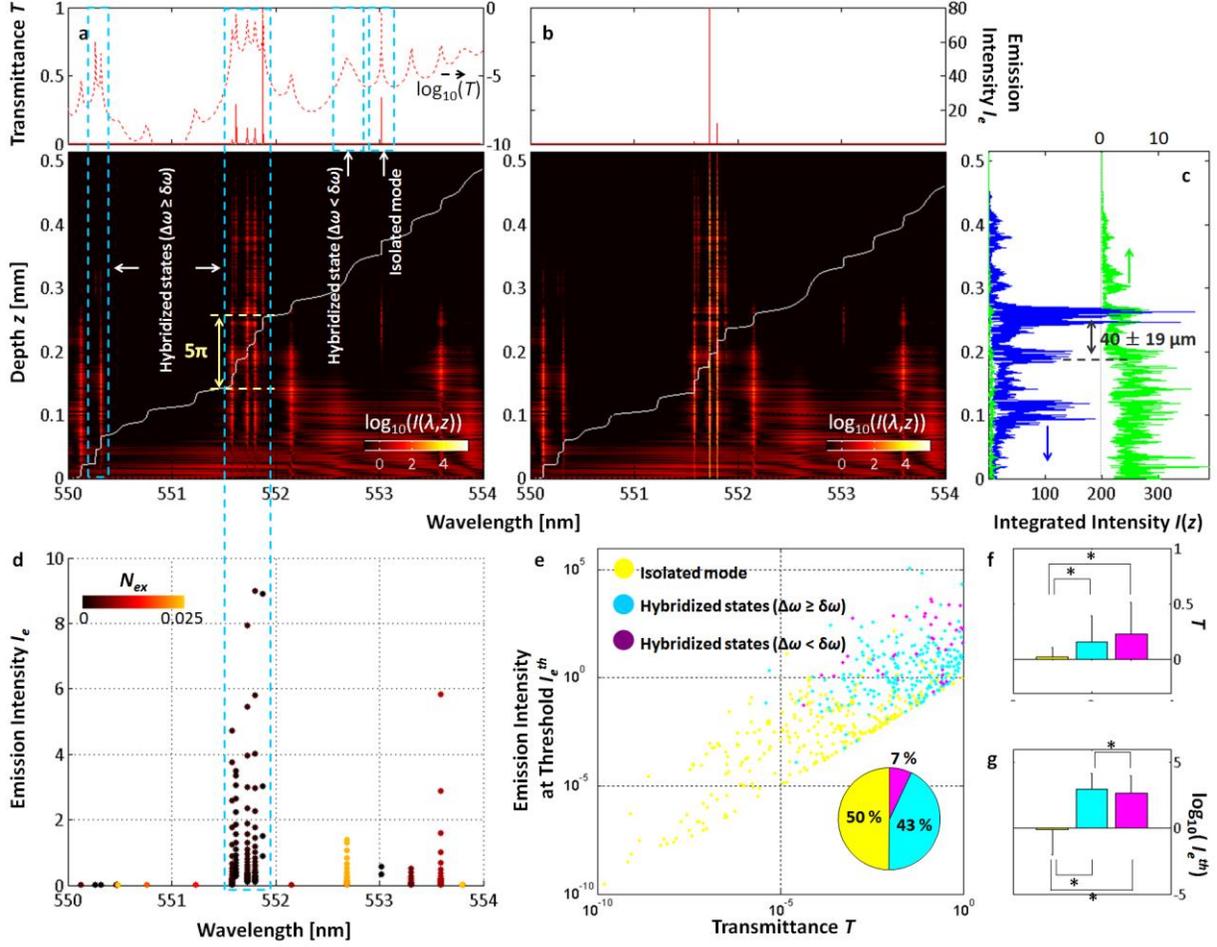

**FIG. 3.** (Color online) Formation of hybridized states and isolated modes in nacre and amplification with gain. (**a**) Transmittance spectra $T$ and $\log_{10}(T)$ (solid and dashed line in upper panel, respectively). Internal field intensity $I$ (lower panel) of quasimodes (without gain). The white solid line is a phase change in the transmittance spectra. (**b**) Emission spectra $I_e$ (upper panel) and internal field intensity $I$ (lower panel) of lasing modes (with gain). (**c**) Spatial profiles of integrated quasimodes (green curve) and lasing modes (blue curve) horizontally shifted for visual clarity. (**d**) $I_e$ of lasing modes as a function of the density of excited atoms $N_{ex}$. (**e**) Emission intensity at the threshold of $I_e^{th}$ (lasing mode) versus $T$ (quasimode) from 32 different disorder configurations. Inset: Occurrence probabilities of two different hybridized states and isolated modes. (**f&g**) Characteristics of $I_e^{th}$ and $T$ of different states and modes. The colors of the data-points in (e) match with the states in the horizontal axes of f&g. * denotes statistically significant differences. The error bars are SDs.

The enhanced transport effect of the hybridized localized states can provide additional benefit for achieving efficient light amplification, because the fields at both excitation and emission bands can overlap spatially deep into the structure.[42] It should be noted that as implied in Fig. 1(e) given the disordered nanostructures, the hybridized states can occur at any wavelength ranges. In contrast to the previous study on isolated localized modes,[42] our results show that in the structures with an extremely large number of layers (> 2,000), resonant tunneling through the isolated localized modes is not as efficient as the hybridized states for light amplification and lasing.

## IV. PHOTOLUMINESCENCE EXPERIMENTS USING NACRE

We conducted lasing experiments in a similar manner as in our previous studies:[43, 44] As shown in Fig. 4(a), a frequency-doubled Q-switched Nd:YAG laser (pulse duration of 400 ps and and $\lambda$ of 532 nm) was used to optically excite each specimen (S in Fig. 4(a)). The excitation beam (diameter = 1 mm) from the pulsed laser was illuminated on the specimen via a low numerical aperture objective (5×). The spot size at the focal plane was 20 µm in diameter and strong focusing onto the specimen surface was avoided. To vary the excitation energy, a linear variable neutral density filter (LF in Fig. 4(a)) was used in the delivery arm. The emitted light on the other side were collected by a fiber bundle through a lens ($L_1$ & $L_2$ in Fig. 4(a)) and a bandpass filter ($BF_1$ in Fig. 4(a)) ($\lambda = 600 \pm 70$ nm) and coupled to a spectrometer (spectral resolution = 0.2 nm). The data acquisition time was 0.2 second, in which 100 excitation pulses were accumulated. In addition, we placed the customized microscopy imaging setup (M in Fig. 4(a)) to image the spatial distribution of emission intensity inside the structures via a bandpass filter of $\lambda = 568 \pm 10$ (FWHM) nm ($BF_2$ in Fig. 4(a)), when the excitation illumination was near the edge. Deproteinized nacre specimens with a thickness of ~ 2 mm were doped with different laser dyes. Four dyes were used at low concentrations of ~ 0.5 mg/ml: Rh6G, rhodamine B (RhB), and rhodamine 101 (Rh101) all in methanol, as well as DCM in dimethyl sulfoxide.

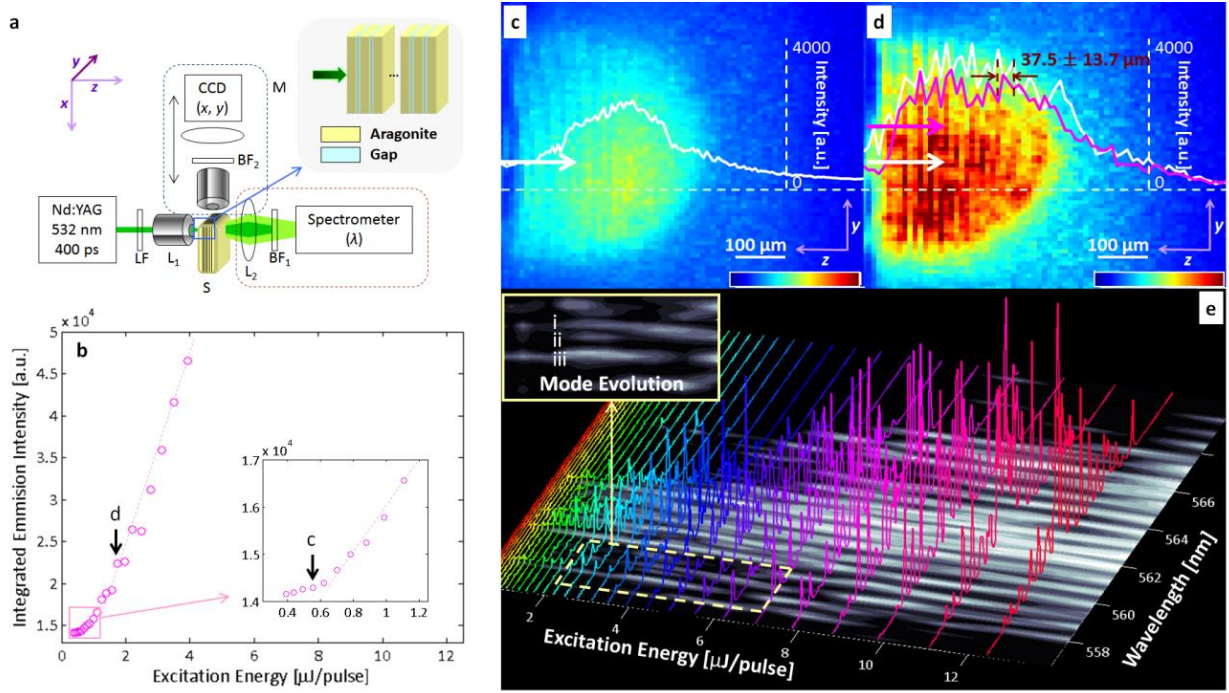

**FIG. 4.** (Color online) Photoluminescence experiments using abalone nacre. (**a**) Schematic diagram of the experimental setup. (**b**) Integrated emission intensity from Rh6G-infilterated nacre as the excitation energy increases. (**c&d**) Pseudocolor microscopic images of spatial fields averaged over wavelength through the 10-nm (FWHM) bandwidth filter: (**c**) Below the threshold (= 0.56 µJ/pulse) and (**d**) above the threshold (= 1.98 µJ/pulse). The solid curves are longitudinal intensity distributions at different horizontal cross-sections. (**e**) Emission spectra as the excitation energy increases. The spectra are also projected onto the grayscale map.

We first observed the signature of hybridized localized states by imaging the spatial distributions of the intensity inside Rh6G-infilterated nacre, when the excitation illumination was perpendicular to the layers near the edge of nacre. Below the threshold (= 0.62 µJ/pulse and Fig. 4(b)), the emission intensity from Rh6G-infilterated nacre is concentrated within ~ 0.5 mm and the spatial distribution is relatively smooth (Fig. 4(c)), because several randomly distributed modes are averaged spatially within the spectral width of the bandpass filter of $\lambda = 568 \pm 10$ (FWHM) nm ($BF_2$ in Fig. 4(a)). However, at an excitation energy of ~7 µJ/pulse (pink circles in Fig. 4(b)), the clear signature of hybridized localized states is observed (Fig. 4(d)): The intensity exhibits several peaks inside nacre and stretches through the specimen. Several peaks appear along the longitudinal direction with a separation of ~ 37.5 ± 13.7 (SD) µm. This fairly consistent separation of intensity peaks indicates that the hybridized states are selectively distinguished via lasing action. Indeed, this inter-resonance spatial separation obtained from the experimental visualization of the local energy disposition in nacre is in agreement with the numerical result (Fig. 3(c)).

As the excitation energy increases, spectral evolution is clearly seen in the grayscale intensity map of Fig. 4(e) and the inset. Multiple narrow emission peaks rapidly emerge from Rh6G-infilterated nacre as the excitation energy increases. Immediately above the threshold, new lasing peaks appear (e.g. modes i, ii, and iii) and then some initial peaks disappear (e.g. mode iii). At higher excitation energy (> 9 µJ/pulse), the spectral positions of the multiple peaks become stabilized. Indeed, such dynamical spectral evolution of mode competition/suppression as a function of the excitation energy has been studied mainly in theoretical

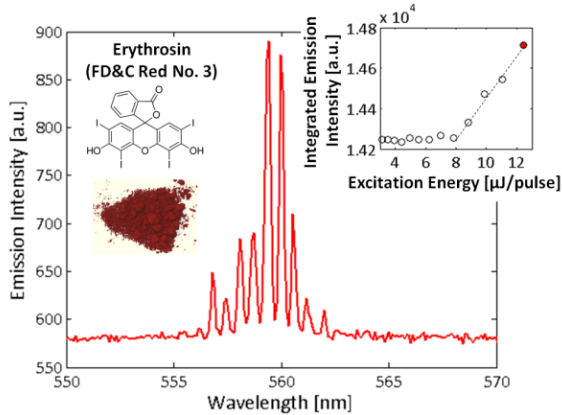

**FIG. 5.** (Color online) Nacre lasers using erythrosine. Representative multimode emission spectrum above the threshold (excitation energy = 12.4 μJ/pulse as indicated the red dot in the right inset). Left inset: Structural formula of erythrosine and photography of erythrosine powders. Right inset: Emission intensity integrated within $\lambda = 547 - 572$ nm as a function of the excitation energy.

studies,[45, 46] due to the inefficient nature of high-dimensional disordered resonators.

To further support the efficient nacre resonator, we utilized erythrosine (FD&C Red No. 3) that is approved by the FDA for food coloring. We measured quantum yields (QYs) of fluorophores, including erythrosine, in a similar manner as in our previous study.[47] The QY of erythrosine in methanol (= 2.5 %) is 37.5 times lower than those of typical laser dyes (e.g. Rh6G = 94.5 %). Surprisingly, nacre still allowed us to achieve multimode lasing action using erythrosine (Fig. 5 and the threshold ~ 8 μJ/pulse). This result supports the idea that nacre provides a means to amplify extremely weak fluorescence light to readily detectable sharp stimulated emission peaks.

Finally, we investigated gain competition and saturation, which can easily be manifested in efficient resonators. We applied excitation energy (15 – 20 μJ/pulse) higher than the lasing threshold and quantified the spectral spacing of discrete emission peaks. Representative multiple narrow emission peaks are shown in Fig. 6(a). Interestingly, Fig. 6(b) reveals that the frequency spacing $\Delta\omega$ (in the unit of cm$^{-1}$) has a highly consistent value. This unique spectral characteristic of multimode lasing is attributable to high competition and saturation of gain.[45, 46, 48, 49] As nacre has low dimensionality and inherently forms hybridized localized states, which are forced to share the same gain spatially, the local depletion of excited electrons from the lasing modes suppresses other spatially overlapped modes. The importance of hole burning in this characteristic is supported by the fact that $\Delta\omega$ from the nacre resonators is comparable to the homogenous linewidths of similar dye molecules measured at room temperature.[50] The highly stable and repeatable multimode lasing emission in nacre could potentially be considered as a precursor of spontaneous mode-locking, as demonstrated by exciting a

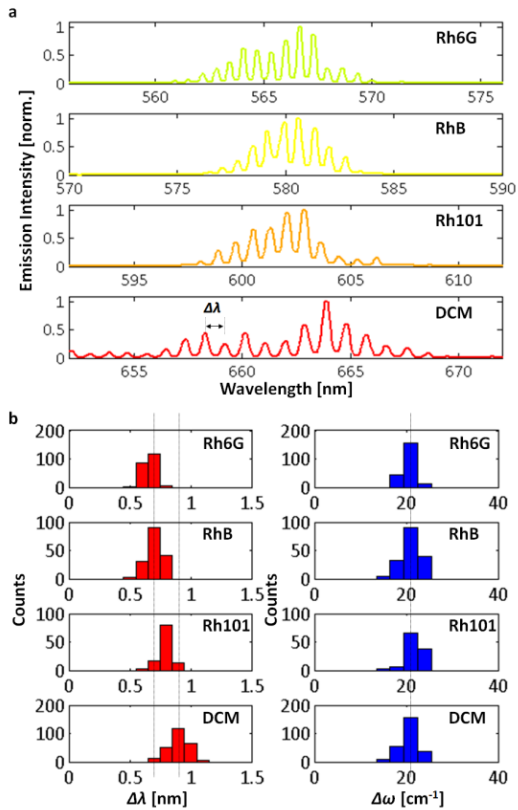

**FIG. 6.** (Color online) Spectral characteristics of lasing emission in nacre originating from hybridized multiple resonances at relatively low excitation energy. (a) Emission spectra at the excitation energy of 15 – 20 μJ/pulse above the lasing thresholds. (b) Wavelength spacing $\Delta\lambda$ and frequency spacing $\Delta\omega$ between adjacent peaks, respectively.

large number of spatially overlapped resonances in disordered cavities.[51]

## V. CONCLUSION

We report that hybridized/coupled multiple resonances in nacre allow collective excitation to synergistically enhance light amplification and lasing. Our study provides a foundation for natural/synthesized nacre materials that can be immediately exploited for developing economical photonic systems and for better understanding wave interactions in complex media. In particular, nanoscale gaps in nacre could potentially be utilized to host/trap other nanomaterials for chemo/biosensor development. In addition, nacre can serve as a model system for addressing fundamental questions in low-dimensional systems.

We thank Azriel Genack at the Queens College of the City University of New York for invaluable discussion on necklace/coupled states in random lasers and Vladimir Shalaev at Purdue University for his insightful commentary. This work was supported in part by the grants from Abbott Laboratories and NIH (R21 ES020965).